# Ferro-type order of magneto-electric quadrupoles in the pseudo-gap phase of $YBa_2Cu_3O_{6+x}$ unearthed by neutron diffraction

S. W. Lovesey[1,2], D. D. Khalyavin[1] and U. Staub[3]

[1]ISIS Facility, STFC Oxfordshire OX11 0QX, UK

[2]Diamond Light Source Ltd, Oxfordshire OX11 0DE, UK

[3]Swiss Light Source, Paul Scherrer Institut, CH 5232 Villigen PSI, Switzerland

A theory that fully describes previously observed neutron Bragg spots indicative of magnetic order in the pseudo-gap phase of $YBa_2Cu_3O_{6+x}$ is presented. Magnetic diffraction intensities, unveiled by polarization analysis, are explained by parity-violating magnetic (magneto-electric) quadrupoles at in-plane Cu sites arranged in a simple ferro-type motif. An absence of conventional spin and orbital angular momentum dipoles in the theory is compatible with the occurrence of superconductivity and strongly suggests, moreover, that magneto-electric quadrupoles are the primary order-parameter in the pseudo-gap phase.

The understanding of superconductivity in cuprates remains to be a challenging problem in condensed matter research. A general phase diagram of cuprates starts with an undoped antiferromagnetic parent compound that transforms when doped to a regime, in which superconductivity appears (underdoped regime), reaching further an optimally doped region with maximum $T_c$. [1] In the underdoped regime, a pseudo-gap phase appears, manifested by modifications of the electronic properties [2] that are observed below a temperature defined as $T^*$. If this temperature is associated with a real phase transition or if it is just a regime of turnover to a precursor state of superconductivity, remains to be settled. In this phase, a wave-vector dependent opening of a (pseudo) gap at the Fermi surface has been observed by angle resolved photoemission. [3] No clear structural modifications have been found at the onset of $T^*$ nor is there a distinct jump observed in specific heat measurements. Measurements of non-reciprocal Faraday rotation imply that cuprates are actually polar, but the observed rotation is extremely small.[4, 5]

$YBa_2Cu_3O_{6+x}$ (YBCO) is the first high-$T_c$ superconductor with transition temperatures above that of boiling liquid nitrogen and crystallizes in the orthorhombic Pmmm-type symmetry. [6] The key structural features are the two $CuO_2$ planes (see Figure 1). Electronic properties of YBCO are almost 2-dimensional and display strong angular anisotropy.[1] A gap associated with an insulating phase only exists for electrons travelling parallel to Cu-O bonds, whilst electrons travelling at 45° to this bond can move freely throughout the crystal. Low energy electrons reside in a single band of carriers formed by hybridization of $Cu^{2+}$ orbitals and oxygen 2p-electrons, with the degree of hybridization essentially controlled by spin-orbit coupling and parity-odd contributions to the electronic potential at a Cu site. An ordering wave vector consistent with a doubling of the in-plane chemical unit-cell describes the antiferromagnetic order in the $CuO_2$ planes of the parent compound. [7]

This order disappears very fast when the material is doped. In the underdoped regime, very strong evidence for the occurrence of an additional second order magnetic phase transition is obtained through polarized neutron diffraction experiments. [8] The onset temperature of magnetic scattering matches approximately $T^*$ of the pseudo gap phase. Later similar results were found on several other cuprate systems. [9] These signals appearing at $T^*$ have been interpreted in the case of YBCO as caused by magnetic dipole moments of $\approx 0.1$ $\mu_B$ inclined at 45º to the c-axis and is discussed in detail in ref. [9]. The observed magnetic order contrasts with the fact that substitutional magnetic impurities at Cu sites destroy superconductivity, which appears in the pseudo-gap phase at low temperatures. The magnetic scattering signals have been discussed [9] in relation to models of orbital currents,[10] and the data were also put in context [9] to data of NMR and µSR, that did not find magnetic signals related to $T^*$. There is, however, an intrinsic problem with the interpretation of the neutron data in terms of scattering by simple magnetic moments. [9] No discernible intensity in the (0, 0, 2) Bragg spot conflicts with data gathered at other Bragg spots, because it implies the absence of an in-plane component of the magnetic dipole moment.

Strong constraint in the interpretation of the observed neutron scattering signals is based on the key observation, that magnetic intensity occurs solely on reflections that are allowed by the chemical structure of the material, which is contrast to the antiferromagnetic [11] order of the parent compound. As there is no overall magnetization observed below $T^*$, it significantly limits the possible models describing the neutron diffraction data based on the scattering on spin and orbital magnetic moments in this material.

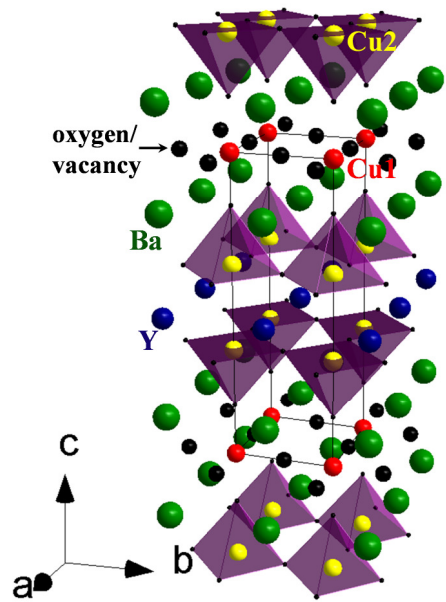

Figure 1 Crystal structure of YBaCu$_3$O$_7$ with the magnetic Cu2 ions in the CuO$_2$ planes.

By exploiting the recent finding that additional terms can appear in the neutron scattering cross-section when local (site) inversion symmetry is absent, [12] we derive a unit-cell structure factor for Bragg diffraction consistent with observed intensities and circumvent the problem of an in-plane magnetic moments. We use operators that are simultaneously magnetic and electric (time and space inversion symmetry broken), including magnetic charge and an anapole (toroidal dipole moment). These magneto-electric multipoles are located at Cu sites that do not have inversion symmetry. The lack of inversion symmetry allows significant admixtures of 2p(O) and 3d(Cu) orbitals involved in bonding and 3d(Cu) and 4p(Cu) orbitals, states with different orbital angular momentum. These orbital overlaps determine the size of magneto-electric multipoles.[13, 14] Magneto-electric multipoles have not only been used recently to interpret resonant x-ray diffraction and absorption data [15, 16], but also have been found in DFT simulations of magneto-electric materials. [14] A beauty of our theory is that a simple ferro-type order of magneto-electric quadrupoles alone are able to fully describe all observed and absent magnetic neutron intensities - giving strong evidence that they can act as an order parameter of a phase with distinct electronic properties, such as the pseudo-gap phase of cuprates.

We base our simulations on a magnetic structure that belongs to the point group 2/m', which is neither polar nor chiral, but allows magneto-electric multipoles at in-plane Cu sites. The absence of ferromagnetism in this system restricts any magnetic order to be antiferromagnetic between the individual $CuO_2$ planes. Considering the amplitude for magnetic neutron scattering,

$$\langle \mathbf{Q} \rangle = \langle \exp(i\mathbf{R} \cdot \mathbf{k}) [\mathbf{S} - i(\mathbf{k} \times \mathbf{p})/\hbar k^2] \rangle, \qquad (1)$$

where $\mathbf{R}$, $\mathbf{S}$ and $\mathbf{p}$ are electron position, spin and linear momentum operators, respectively, [17] we calculate Bragg diffraction at a scattering wavevector $\mathbf{k} \equiv (h, k, l)$ with Miller indices h, k, l. The scattering amplitude is $\langle \mathbf{Q}_\perp \rangle = \boldsymbol{\kappa} \times (\langle \mathbf{Q} \rangle \times \boldsymbol{\kappa})$ using a unit vector $\boldsymbol{\kappa} = \mathbf{k}/k$. In these expressions, angular brackets $\langle \ ... \ \rangle$ denote the expectation value, or time-average, of the enclosed operator.

In general, neutrons are deflected by parity-even and parity-odd multipoles that are time-odd (magnetic).[12, 17] On treating $\mathbf{k}$ as a small quantity in (1), $\langle \mathbf{Q} \rangle$ contains a spin moment $\langle \mathbf{S} \rangle$ at the initial level of an expansion. Some algebra is required to show that, $\mathbf{R}$ and $\mathbf{p}$ in (1) produce a contribution proportional to orbital angular momentum, $\mathbf{L} = \mathbf{R} \times \mathbf{p}$, in the second level of an expansion.[17, 18] Combining the two results, we may write

$$\langle \mathbf{Q} \rangle \approx (1/2)\, f(k)\, \langle \mathbf{L} + 2\mathbf{S} \rangle \qquad (2)$$

in which f(k) is a normalized atomic form factor with f(0) = 1. This well-established, small **k** limit for ⟨**Q**⟩ is routinely exploited to determine size and direction of ordered magnetic dipoles, ⟨**L** + 2**S**⟩. At the second level in an expansion in **k**, the spin contribution to ⟨**Q**⟩ contains i⟨(**k**•**R**)**S**⟩ ≡ (ikR/2)[**κ** x ⟨**S** x **n**⟩ + ⟨**S**(**κ**•**n**) + (**κ**•**S**)**n**⟩] in which **n** = **R**/R. Here ⟨**S** x **n**⟩ is the spin anapole, which was studied by Zel'dovich in the course of investigating electromagnetic interactions that violate parity [19, 20]. A magnetic monopole (charge) defined as ⟨**S**•**n**⟩ does not contribute to the scattering amplitude ⟨**Q**$_\perp$⟩ as its contribution from ⟨(**k**•**R**)**S**⟩ is multiplied by **k**. Use of an identity for the angular part of linear momentum reveals an orbital anapole in ⟨**Q**⟩ in addition; the identity, or operator equivalent, is **p**$_\omega$ ≡ (1/2R) **Ω** with **Ω** = **L** x **n** − **n** x **L** an orbital anapole.[21] In the presence of a centre of inversion symmetry, ⟨**Ω**⟩ = ⟨**S** x **n**⟩ = 0 and ⟨**Q**$_\perp$⟩ regains its standard form of Eq. 2.

Central to simulations of a diffraction pattern is an electronic structure factor [22],

$$\Psi_{K,Q} = \sum_\mathbf{d} \exp(i\mathbf{d} \cdot \mathbf{k}) \langle U^K_Q \rangle_\mathbf{d}. \qquad (3)$$

Here, the vector **d** labels the magnetic Cu sites and ⟨$U^K_Q$⟩ are spherical atomic multipoles of rank K with projections − K ≤ Q ≤ K that are constrained by elements of symmetry in the site point-group. These magnetic multipoles are either parity-even (⟨$T^K_Q$⟩) or parity-odd (⟨$H^K_Q$⟩ and ⟨$O^K_Q$⟩) [12] with familiar magnetic dipole ⟨**T**$^1$⟩ ∝ ⟨**L** + 2**S**⟩, spin anapole ⟨**H**$^1$⟩ ∝ ⟨**S** x **n**⟩ and orbital anapole ⟨**O**$^1$⟩ ∝ ⟨**Ω**⟩.

Expressions for $\Psi_{K,Q}$ and the scattering amplitude ⟨**Q**⟩ are presented in Supplementary Material [23] for two candidate magnetic structures, namely, P2/m' and C2/m' that are subgroups of P4/mmm1'. SM also contains details of our interpretation of data gathered using neutron polarization analysis at the Bragg spot indexed by (1, 0, 1) that is reproduced in Figure 2 [9]. An absence of magnetic intensity at (0, 0, 2) is shown by us to imply that allowed anapoles and the in-plane magnetic moment are absent. [9] Moreover, the magnetic space-group P2/m' does not allow spin-flip scattering (SF) for all three configurations of the neutron polarization at the (1, 0, 1) reflection. [23] Since this finding contradicts data displayed in Figure 2 the candidate P2/m' can be discarded. [9] On the other hand, there is no such contradiction with magnetic space-group C2/m' for which,

$$(\langle \mathbf{Q}_y \rangle - \langle \mathbf{Q}_x \rangle) \approx -12i \sqrt{(1/5)} \cos(\varphi) \kappa_x \langle H^2_{+2} \rangle',$$

$$(\langle \mathbf{Q}_y \rangle + \langle \mathbf{Q}_x \rangle) \approx -2i \sqrt{(6/5)} \cos(\varphi) \kappa_x \langle H^2_0 \rangle,$$

$$\langle \mathbf{Q}_z \rangle \approx 3i \sin(\varphi) \langle T^1_z \rangle + 2i \sqrt{(6/5)} \cos(\varphi) \kappa_z \langle H^2_0 \rangle, \qquad (4)$$

at the Bragg spot (1, 0, 1). Orthogonal (x, y, z) are derived from {(1, −1, 0), (1, 1, 0), (0, 0, 1)} for C2/m', in which Cu(2) ions use sites 4i that have symmetry m'. Magneto-electric quadrupoles $\langle H^2_{+2} \rangle'$ and $\langle H^2_0 \rangle$ in (4) are manifestations of a state of magnetic charge, while $\langle T^1_z \rangle \propto \langle (\mathbf{L} + 2\mathbf{S})_z \rangle$ and the z-axis coincides with the crystal c-axis. At this juncture one should re-evaluate the symmetry of the material, taking account only multipoles that survive in (4). Results (4) actually flow from the subgroup Cm'm'm' with Cu(2) using sites 4k which have symmetry m'm'2. Additional constraints in this site symmetry, compared to m', forbid all multipoles that contribute to the intensity of the Bragg spot (0, 0, 2), namely, anapoles, quadrupoles with odd projections, and in-plane magnetic dipoles. [23] Crystal classes 2/m' (P2/m' & C2/m') and m'm'm' (Cm'm'm') are neither polar nor chiral.

Notably, our solution (4) does not require $\langle (\mathbf{L} + 2\mathbf{S})_z \rangle$ to be non-zero to obtain $\langle \mathbf{Q}_z \rangle \neq 0$ and naturally resolves the previous problem with the in-plane magnetic moments. In consequence, a state of magnetic charge is sufficient to fully account for available neutron diffraction data. For this case, a ratio of spin-flip intensities,

$$SF(c)/SF(b) = (\langle H^2_0 \rangle / \langle H^2_{+2} \rangle')^2 (1/6) (\kappa_c + \kappa_a (a/c)\sqrt{2})^2, \qquad (5)$$

demonstrates a direct measure of the ratio of magneto-electric quadrupoles sketched in figure 3. From data reproduced in Figure 2 we find SF(c)/SF(b) ≈ 0.9 ± 0.2 yielding $(\langle H^2_0 \rangle / \langle H^2_{+2} \rangle')^2$ ≈ 9.5 ± 2 that is independent of temperature in the pseudo-gap phase, to a good approximation. The three sets of data in Figure 2 are not independent in view of a polarization sum-rule SF(a) = SF(b) + SF(c). [9, 23]

Note that there is no direct relation between our theory, using magnetic charge at Cu sites, and loops of current in $CuO_2$ planes. [10, 11] Such loops create a magnetic dipole moment from orbital angular momentum that is off-set from Cu sites, and this construct does not explain the neutron diffraction under discussion. [9] In a recent theoretical study,[24] the quantum model of a magnetic charge coupled to a quantum rotator by Stone [25] has been adapted to incorporate a crystal field, so that it describes all magneto-electric multipoles appearing in (4).

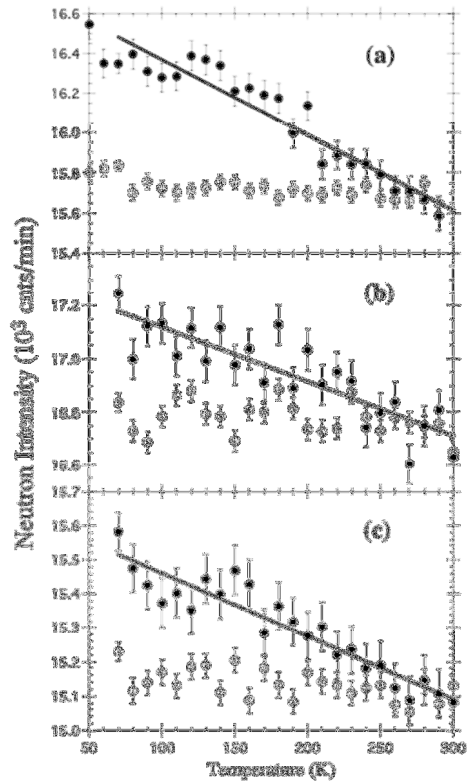

Figure 2. Spin-flip (SF) intensities of the Bragg spot **k** = (1, 0, 1) from YBCO as a function of temperature together with equation (5) (solid line) for the ratio of intensities. Labelling (a), (b) and (c) of panels is used in Supplementary Material [23] that contains details of the confrontation of theory and experimental data, including the polarization sum-rule. In panel

(a) neutron polarization **P** and wavevector **k** are parallel. On the other hand, **P** and **k** are orthogonal in panels (b) and (c), with **P** parallel to the b-axis in (b) and **P** in the a-c plane in (c). Polarization analysis is employed in diffraction to separate magnetic from nuclear signals that are superimposed because the magnetic motif coincides with chemical structure. Data are reproduced from Fig. 5 of ref. [9].

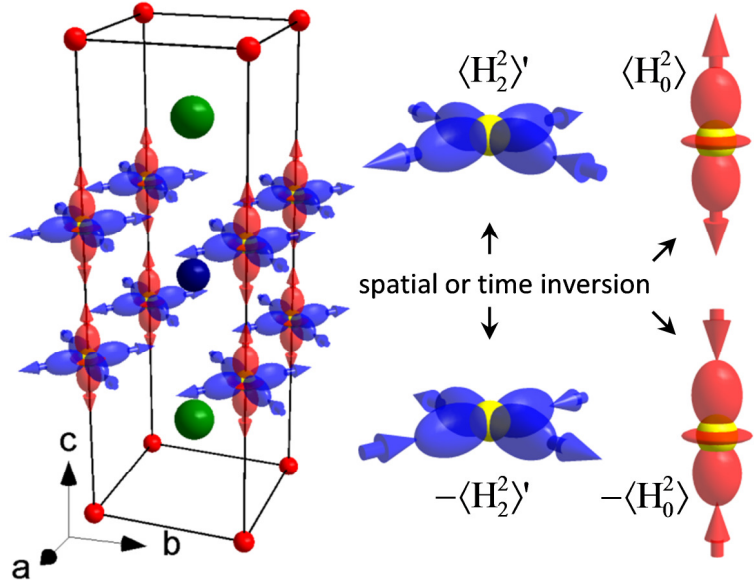

Figure 3. Ferro-type ordering of magneto-electric quadrupoles in $CuO_2$ planes for YBCO in the pseudo-gap phase. Arrows indicate spin directions in magnetic charge-like quadrupoles $\langle H^2_0 \rangle \propto \langle 3S_z n_z - \mathbf{S} \cdot \mathbf{n} \rangle$ and $\langle H^2_{+2} \rangle' \propto \langle S_x n_x - S_y n_y \rangle$, together with their response to spatial or time inversion.

In summary, we present a new direction in understanding electronic properties of the pseudo-gap phase of cuprates with a theory that fully describes unexpected magnetic Bragg spots in the YBCO neutron diffraction pattern. [9] Scattering by magneto-electric multipoles, originating from magnetic charge, removes the intrinsic inconsistency in an interpretation using conventional magnetic dipoles. Encouragingly, our theory of magneto-electric quadrupoles arranged in a ferro-type motif is very simple. Quadrupoles localized on Cu ions using sites 4k in the magnetic space-group Cm'm'm' act as the order-parameter for the pseudo-gap phase of YBCO. The corresponding crystal class (point group) m'm'm' allows the Kerr effect that has already been reported for the material. [4, 26, 27]

Acknowledgements. We thank P. Bourges for comments on an early draft of the communication and making data for Figure 2 available to us. One of us (US) is grateful to C. Niedermayer and M. Ramakrishnan for useful discussions. This work was in part supported by the Swiss National Science Foundation.